\documentclass[twocolumn,amsmath,amssymb,prd]{revtex4}

\usepackage{graphicx}
\usepackage{rotating}
\def\arcdeg{$^{\circ}$}

\begin{document}

\title{Detailed Abundances for a Large Sample of Giant Stars in the Globular Cluster 47 Tucanae (NGC 104)}
\author{M.~J. Cordero and C. A. Pilachowski}
\affiliation{Astronomy Department, Indiana University Bloomington, Swain West 319, 727 E. 3rd Street, Bloomington, IN 47405-7105, USA; majocord@indiana.edu; catyp@astro.indiana.edu}

\author{C.~I. Johnson\footnote{Visiting Astronomer, Cerro Tololo Inter-American Observatory.  CTIO is operated by AURA, Inc.\ under contract to the National Science Foundation; National Science Foundation Astronomy and Astrophysics Postdoctoral Fellow; Clay Fellow.}}
\affiliation{Harvard-Smithsonian Center for Astrophysics, 60 Garden Street, MS-15, Cambridge, MA 02138 USA; cjohnson@cfa.harvard.edu}

\author{I. McDonald and A. A. Zijlstra}
\affiliation{Jodrell Bank Centre for Astrophysics, Alan Turing Building, Manchester M13 9PL, UK; mcdonald@jb.man.ac.uk; albert.zijlstra@manchester.ac.uk}

\author{J. Simmerer\footnote{Visiting Astronomer, Cerro Tololo Inter-American Observatory.  CTIO is operated by AURA, Inc.\ under contract to the National Science Foundation} }
\affiliation{University of Utah, Physics and Astronomy, 115 South 1400
East \#201, Salt Lake City, UT 84112-0830 USA;
jennifer@physics.utah.edu}

\author{Accepted for publication in The Astrophysical Journal}

\begin{abstract}
47 Tuc is an ideal target to study chemical evolution and GC formation in massive more metal-rich GCs since is the closest, massive GC. We present chemical abundances for O, Na, Al, Si, Ca, Ti, Fe, Ni, La, and Eu in 164 red giant branch (RGB) stars in the massive globular cluster 47 Tuc using spectra obtained with both the Hydra multi-fiber spectrograph at the Blanco 4-m telescope and the FLAMES multi-object spectrograph at the Very Large Telescope. We find an average [Fe/H]=--0.79$\pm$0.09 dex, consistent with literature values, as well as over-abundances of alpha-elements ($[\alpha/\mbox{Fe}]\sim0.3$ dex). The n-capture process elements indicate that 47 Tuc is r-process dominated ([Eu/La]=+0.24), and the light elements O, Na, and Al exhibit star-to-star variations. The Na-O anti-correlation, a signature typically seen in Galactic globular clusters, is present in 47 Tuc, and extends to include a small number of stars with [O/Fe] $\sim$\,--0.5. Additionally, the [O/Na] ratios of our sample reveal that the cluster stars can be separated into three distinct populations.  A KS-test demonstrates that the O-poor/Na-rich stars are more centrally concentrated than the O-rich/Na-poor stars.  The observed number and radial distribution of 47 Tuc's stellar populations, as distinguished by their light element composition, agrees closely with the results obtained from photometric data. We do not find evidence supporting a strong Na-Al correlation in 47 Tuc, which is consistent with current models of AGB nucleosynthesis yields.
\end{abstract}


\maketitle

\section{Introduction}

Recent observational evidence has shown that many Galactic globular clusters, including 47 Tucanae (NGC 104), contain more than one chemically distinct stellar population (e.g., see recent review by Gratton et al. 2012).  In the case of 47 Tucanae ([Fe/H] $\sim$--0.77 from Carretta et al. 2009a; age 9.9 $\pm$ 0.7 Gyr from Hanson et al. 2013, although VandenBerg et al. reported an age of 11.75 $\pm$ 0.25 Gyr), Anderson et al. (2009) used archival Hubble Space Telescope images to resolve at least two populations on the cluster's subgiant branch.  Additional analysis of the cluster's narrow-red-horizontal and subgiant branch regions by Di Criscienzo et al. (2010) identified three subpopulations: a first generation population comprising 30\% of cluster stars and two second generation populations.  The dominant population of second generation stars are thought to have enhanced He, but the minority population ($\sim$10\%) may instead exhibit enhancements in the sum of C+N+O.  Nataf et al. (2011) also find changes in the RGB-bump and horizontal branch morphologies as a function of cluster radius to be evidence supporting He-enhancement in some 47 Tuc stars near the core.

More recently, Milone et al. (2012) extended these studies to follow the two major populations in 47 Tuc from the main sequence through the horizontal branch, and identified a third population comprising only 8\% of the cluster's stars and discerned only on the subgiant branch.  The primordial population contributes only 20\% of the stars in the cluster center, but is more extended, and contributes 30\% of the cluster's stars overall. In addition, Richer et al. (2013) reported that stars on the blue side of the main sequence are more centrally concentrated and show anisotropic proper motions compared to stars on the red side of the main sequence.  Overall, the evidence for multiple populations in 47 Tuc is compelling and suggests that both chemical and kinematic differences separate the populations.

Studies of the light elements (CNO, Na) in 47 Tuc mirror the photometric analyses of the cluster's multiple populations.  Worley \& Cottrell (2012), following earlier analyses by Norris \& Freeman (1979) and Paltoglou \& Freeman (1984), confirmed the anti-correlation of sodium and CN, as well as the bi-modal distribution of CN strength, from medium resolution spectra of giants.  Carretta et al. (2009b) determined oxygen and sodium abundances for 115 giants in 47 Tuc, finding that 2/3 of the stars belong to an intermediate (O-poor and Na-rich) population, consistent with current cluster formation models, with most of the remaining stars belonging to an earlier (primordial) generation.  D'Orazi, et al. (2010) examined several dozen turnoff stars in 47 Tuc, confirming the anti-correlation of oxygen and sodium in unevolved stars and finding a similar distribution in [Na/O] as in the giants.  Their observations rule out in-situ mixing in the giants as the primary source of the O-Na anticorrelation.  Neutron-capture elements, however, show only modest enhancements and no star-to-star variations (James et al. 2004, Worley et al. 2010), suggesting that core-collapse supernovae and low-mass AGB stars also do not cause the light element abundance anomalies.

According to Vesperini et al. (2013) N-body simulations indicate that the multiple populations in a GC do not mix completely until the cluster has lost a large fraction of its mass through two-body interactions, which takes several relaxation times. For instance, the authors found that the populations in a GC are well mixed when the cluster's age is $\sim15$ times its half-mass relaxation time-scale $t_{rh}$; for 47 Tuc the ratio of the cluster age (9.9 Gyr, see Hansen et al. 2013) to its half-mass relaxation time (3.5 Gyr, Harris 1996, 2010 edition) is $\sim3$, thus the multiple populations in 47 Tuc are expected to differ not only on their chemical composition, but also on their radial distribution and proper motions.

In this work we present the analysis of the spectra of 164 red giants in
the globular cluster 47 Tuc, including abundances for oxygen,
sodium, and aluminum, among other elements, to explore star-to-star variations in the light element abundances and characterize the multiple populations. The paper is organized as follows. Sections 2 and 3 present the
observational data and a description of the analysis,
respectively. Our results and a comparison to the literature are
presented in Section 4. In Section 5, we discuss our results in the context of the multiple stellar populations in 
47 Tuc, focusing on the chemical signatures of the multiple populations and their radial extent in the cluster, comparing our findings to theoretical models. Our conclusions are given in section 6.

\section{Observations and Data Reduction}

\subsection{Observations}

Spectroscopic observations were obtained using two multi-fiber positioners and spectrographs: Hydra
on the Blanco 4-m telescope at the Cerro Tololo Interamerican Observatory during two
observing runs in July, 2003, and August, 2010, and FLAMES--GIRAFFE on the VLT--UT2 telescope at the European Southern Observatory on Cerro Paranal during a single observing run in November, 2011. For both Hydra runs the ``large'' (300
$\mu$m) fiber cable, the Blue Air Schmidt Camera, the 316 l/mm
(57.5\arcdeg) echelle grating, and the SITE 4096 $\times$ 2048
pixel CCD were employed, but at two
different wavelength ranges. For the 2003 observations, filter \#6
centered at 6757 \AA\ yielded spectra ranging from $\lambda\lambda$6490-6800 \AA, at a dispersion
of 0.15 \AA\ pixel$^{-1}$. The 2010 observations employed the
E6257 filter, covering the
spectral range from $\lambda\lambda$6100-6350 \AA\ in one
observation, with a resolving power of typically 20,000.\footnote{The E6257 filter is on permanent loan at CTIO and
is available for public use.} Different fiber configurations were
used during each run, although several stars were included in both
configurations. 

Spectra for the VLT--FLAMES portion of the data set utilized the FLAMES--GIRAFFE spectrograph (Pasquini et al. 2003). The high--resolution HR13, HR14A, and HR15 spectrograph setups provided continuous wavelength coverage from approximately 6115--7000 \AA, with a resolving power of R$\approx$20,000 (see also Table 1). 

\subsection{Selection of Stars}
Stars observed with Hydra were selected for observation from the proper motion survey
of Tucholke (1992), with membership probabilities above 88\%. The
proper motion study included stars out to a radius of 32 arc
minutes from the center of the cluster, but does not include stars
within about 2 arc minutes of cluster center. The stars included
in this study range in magnitude from 13.3 $\leq$ B $\leq$ 14.6,
from the tip of the giant branch to just above the horizontal
branch, and include both red giants and asymptotic branch giants.
Stars were selected randomly and not biased as to light element abundance. Spectra of 52 stars were available
from the 2003 observations, and 48 stars were obtained from the
2011 observations, with 17 stars in common between the two
configurations. B magnitudes were taken from Tucholke (1992), and
identifications were matched to either stars in the Lee (1977) or Chun \& Freeman (1978) 
photometric studies to obtain V magnitudes. J and K magnitudes were
obtained from the 2MASS Point Source Catalog. The stars
included in this study are listed in Table 2. The identification numbers without prefix were taken from Lee (1977), while identification numbers with prefix D or E were taken from Chun \& Freeman (1978), and with prefix M were taken from McDonald et al. (2011). 

Stars observed with FLAMES were selected from photometry compiled by McDonald et al. (2011).  The targets were originally selected to contain RGB and early AGB stars of similar temperature and luminosity.  However, a problem with the fiber assignment process led to RGB stars only being observed.  The final sample includes 113 RGB stars from the luminosity level of the horizontal branch to near the RGB-tip.

\begin{figure}
\includegraphics[width=0.5\textwidth]{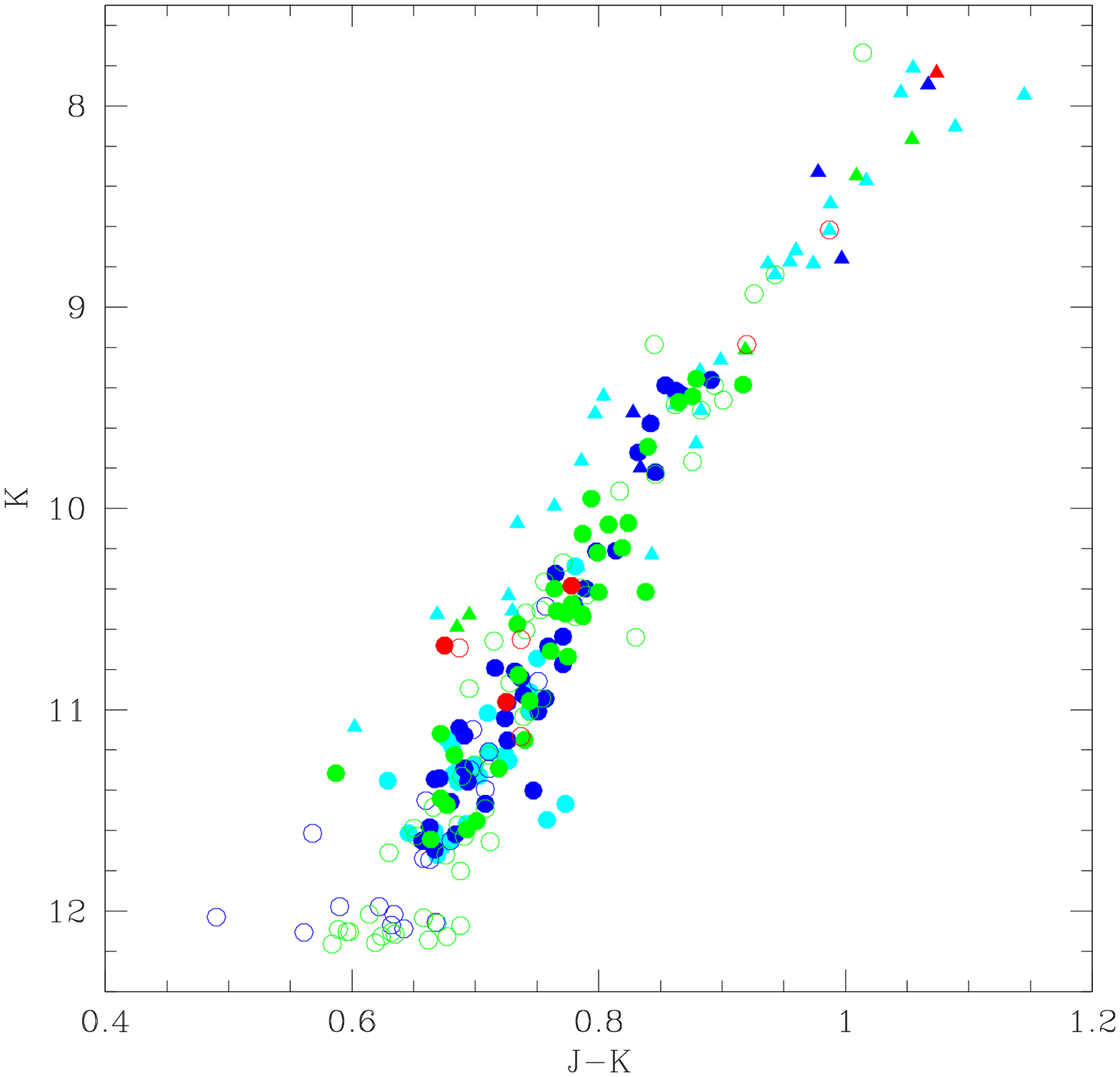} 
\caption{K vs. J-K color magnitude diagram of stars observed in 47 Tuc.  Stars assigned to the primordial population are shown in blue, stars assigned to the intermediate population are shown in green, and stars assigned to the extreme population are shown in red (see section 5.1 for the definitions of population groups).  Our FLAMES and hydra samples are represented by filled circles and triangles, respectively.  Stars from Carretta et al. (2009b) are shown with open symbols.  J and K magnitudes were obtained from the 2MASS Point Source Catalog.}
\label{fig_CMD}
\end{figure}

A color-magnitude diagram of stars observed in 47 Tuc is shown in Fig. \ref{fig_CMD}, including both our Hydra and FLAMES samples and the sample from Carretta et al. (2009b).  Since each of these samples adopted V magnitudes from different sources, we shifted the V magnitudes of the Carretta et al. sample upward by +0.01 magnitudes based on 20 stars in common between the Carretta and FLAMES samples, and shifted the V magnitudes of the Hydra sample downward by -0.05 magnitudes based on 14 stars in common with the FLAMES sample.  Anticipating the assignment of stars to the primordial, intermediate, and extreme populations defined by Carretta et al. (2009b), we have color-coded the stars by population assignment as noted in the figure caption.  Unclassified stars lacking either Na or O abundances are also identified.  The combined sample includes primarily giants on the first ascent of the red giant branch, as well as a few asymptotic giant branch stars.

\subsection{Data Reduction}

Calibration data for both Hydra runs included bias and flat field frames on all
nights. Daylight sky spectra (2003 observations) or comparison lamps (2010 observations) were obtained
for wavelength calibration. Raw images were bias corrected and
trimmed, and an averaged zero frame was subtracted to remove any
two dimensional structure in the bias.  Following these
preliminary reductions, extraction to one dimensional spectra, sky
subtraction, and wavelength calibration were performed using the
IRAF\footnote{IRAF is distributed by the National Optical
Astronomy Observatories, which are operated by the Association of
Universities for Research in Astronomy, Inc., under cooperative
agreement with the National Science Foundation."} script {\it
dohydra}. Finally, the continuum for each spectrum was
normalized using a low order cubic spline. A signal-to-noise
ratio of typically 120-200 per pixel, depending on wavelength,
characterizes the final spectra.

Basic data reduction for the FLAMES--GIRAFFE sample observations was carried out using the Geneva Observatory girBLDRS\footnote{The girBLDRS data reduction package can be downloaded at: http://girbldrs.sourceforge.net/} software.  The reduction tasks included bias subtraction, flat--field correction, wavelength calibration using the ThAr comparison lamps, cosmic ray removal, and object spectrum extraction.  The IRAF task \emph{skysub} was used to create and subtract an average sky spectrum for each exposure, based on 15 sky fibers distributed across the field--of--view.  Additionally, we removed telluric features with the IRAF task \emph{telluric} and a previously obtained grid of rapidly rotating B stars observed at various airmasses.  The individual exposures were then co--added using the IRAF task \emph{scombine} to obtain final spectra with signal--to--noise (S/N) ratios ranging from about 50--200.

\section{Analysis and Uncertainties}
A detailed analysis of the composition of 47 Tuc giants was performed using the 2010
version of MOOG (Sneden 1973).  LTE model atmospheres were
interpolated with the $\alpha$-enhanced ATLAS9 grid of non-overshoot models (Castelli \& Kurucz 2003). Models were interpolated with an adopted
metallicity of [M/H]=--0.75\footnote{We use the standard spectroscopic notation where
[A/B]$\equiv$log(N$_{\rm A}$/N$_{\rm B}$)$_{\rm star}$--log(N$_{\rm A}$/N$_{\rm B}$)$_{\odot}$ and
log $\epsilon$(A)$\equiv$log(N$_{\rm A}$/N$_{\rm H}$)+12.0 
for elements A and B.}.

Initial model atmosphere parameters were estimated using the Alonso et al. (1999) analytic functions relating color to temperature, which were based on the infrared flux method. B-V colors were used for stars for which infrared magnitudes are not
available. Different authors present a range
of values for interstellar reddening to 47 Tuc ranging from
0.029 $\leq$ E(B-V) $\leq$ 0.064 (e.g. Crawford \& Snowden 1975, Gratton et al. 1997). A reddening value of E(B-V)=0.04 (Harris 1996, 2010 edition)
was adopted, in the mid-range of published values and consistent
with the values adopted by most other studies of 47 Tuc's
composition. For stars in the FLAMES sample with large spectral coverage, effective temperatures were constrained via excitation equilibrium.

\begin{figure}
\includegraphics[width=0.5\textwidth]{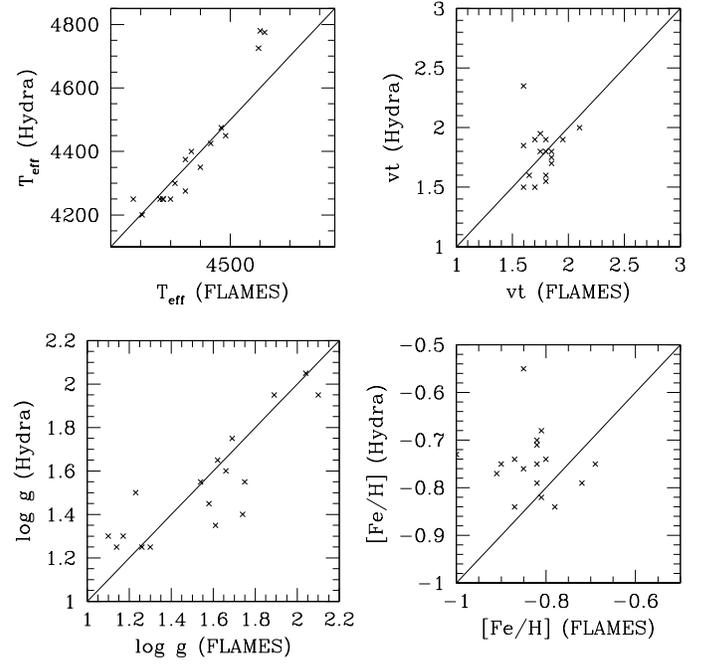} 
\caption{Comparison of our adopted model atmosphere parameters for Hydra and FLAMES observations. The diagonal line indicates perfect agreement.}
\label{fig_logTeff}
\end{figure}

Surface gravities were obtained using the absolute
bolometric magnitudes, derived effective temperatures, and assuming a mass of 0.8 $M_{\odot}$ for all stars in the Hydra sample. Bolometric corrections were adopted from Alonso et al. (1999), and we used a distance modulus
of (m-M)$_{V}$ = 13.37 (Harris 1996, 2010 edition). The limited spectral coverage
available with the Hydra multi-object spectographs provides too
few lines of Fe II to permit a reliable spectroscopic estimate of
surface gravity. For stars in the FLAMES sample with wider spectral coverage, surface gravities could be obtained from the ionization equilibrium of Fe I and Fe II. Microturbulence was estimated for each star using
the v$_{t}$ relation given by Johnson et al. (2008), and then
adjusted to minimize the dependence of derived [Fe/H] on
line strength. The final adopted atmospheric
parameters are included in Table 3. A comparison of the atmosphere parameters obtained for a subsample of 17 stars using both Hydra and FLAMES data is shown in Fig. \ref{fig_logTeff}. The average differences in the sense FLAMES minus Hydra are $\Delta(T_{\mbox{eff}})=-18$ K ($\sigma=75$ K), $\Delta(\log g)=0.02$ cgs ($\sigma=0.16$ cgs), $\Delta([Fe/H])=-0.08$ dex ($\sigma=0.10$ dex), and $\Delta(v_t)=-0.02$ km s$^{-1}$ ($\sigma=0.24$ km s$^{-1}$). The average difference for all [x/Fe] ratios are smaller than 0.09 dex and are shown in column 9 of Table 3. The [x/Fe] abundance ratios of the Hydra sample shown in Table 2 were corrected by these offsets. Thus, no strong systematic offsets are introduced by our adopted atmosphere parameters in our total sample.

For iron, nickel, silicon, titanium, and calcium, abundances were
determined from equivalent widths, adopting the line list from Johnson \& Pilachowski (2010). As described by
Johnson \& Pilachowski (2010), gf values were determined by matching line
strengths in the solar spectrum to the photospheric abundances of
Anders and Grevesse (1989). While the Anders and Grevesse
photospheric abundances have been supplanted by later studies, we
have retained their use to assure that our abundances are on the
same scale as those of Johnson et al. (2009) and subsequent
papers.

Equivalent widths for the Hydra samples were measured using the \textit{splot} task in
IRAF. Only lines with log W/$\lambda$ $<$--4.5 were
included in our analysis. Similarly, for the FLAMES data the equivalent widths were measured using an updated version of the code developed for Johnson et al. (2008).  Isolated lines were fit with a single Gaussian profile.  Partially contaminated lines were deblended by fitting multiple Gaussian profiles and guided by an examination of the high resolution Arcturus atlas by Hinkle et al. (2000).

Abundances of the elements oxygen, sodium, aluminum, lanthanum,
and europium were determined using spectrum synthesis to account
for blending with other spectral features. Synthetic spectra were
calculated for each spectral region using each stars' adopted
model atmosphere parameters and a range of abundances for the
target species. Synthetic spectra were broadened by a Gaussian smoothing kernel to
match the observed line profiles of nearby, unblended lines and
compared by eye to the observed stellar spectra to determine the
best-fit abundance.  Uncertainties of abundances determined using
spectrum synthesis are set from the smallest clear distinction
that can be made consistent with the S/N ratio of the observed
spectrum.

\begin{figure}
\includegraphics[width=0.5\textwidth]{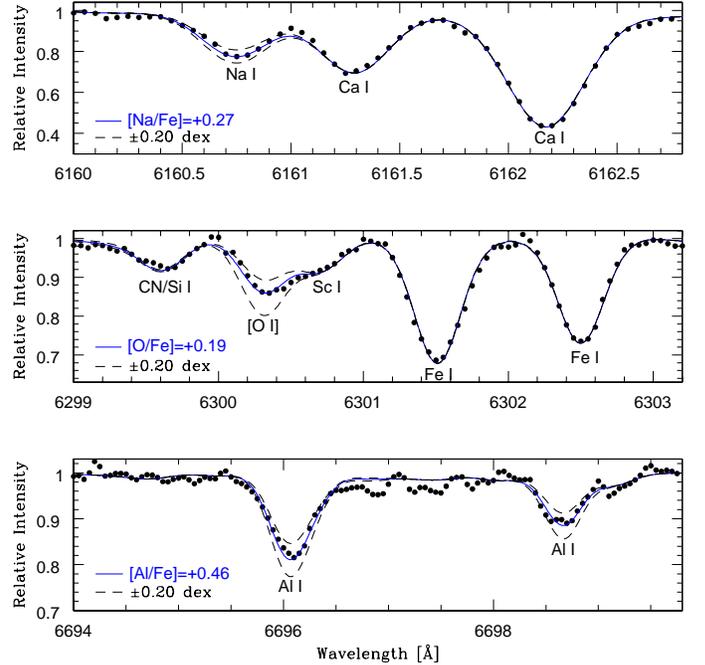} 
\caption{Sample spectrum of D 229 showing the determination of the sodium, oxygen, and aluminum abundances using spectrum synthesis.  The observed spectrum is shown as dots, while the synthetic spectra are shown as dashed lines for three different abundances.  The adopted abundance is shown as a solid blue line.  Differences of 0.2 dex can be easily discerned.}
\label{fig_synth}
\end{figure}

\begin{figure}
\includegraphics[width=0.5\textwidth]{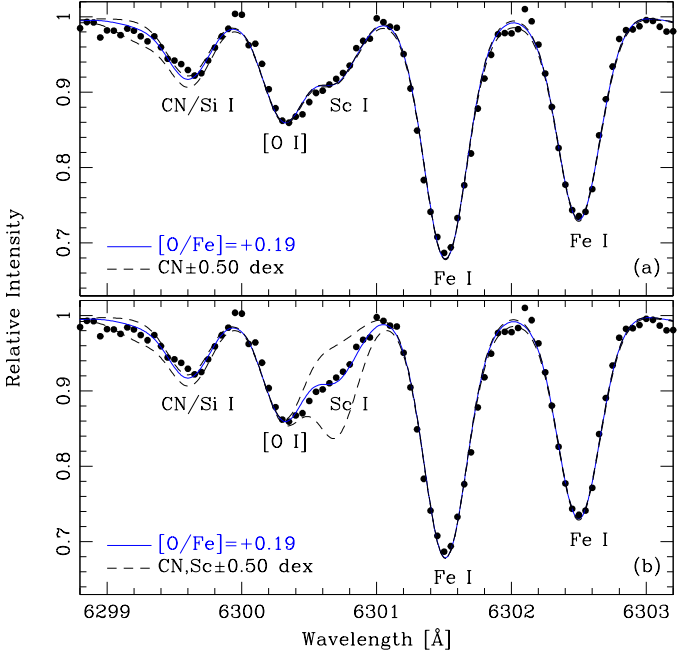} 
\caption{Spectrum synthesis around the oxygen line showing the effect of altering the abundances of blended features. The upper panel shows that the oxygen abundance is unaffected by a change in CN abundance of $\pm$ 0.50 dex (dashed lines).  The lower panel shows that the Sc feature must be included in the synthesis to derive a reliable oxygen abundance.}
\label{fig_CNSc}
\end{figure}

The oxygen abundance was determined using the [O I] $\lambda$6300
\AA\ feature. The line list was the same as that used in Johnson
\& Pilachowski (2010) and the synthesized spectral region covered
6296-6304 \AA. Since the O[I] line is moderately sensitive to
the C+N abundance, for each star, the synthetic spectrum was
obtained by modifying the C and N abundances until a good match
was found between the synthetic and observed spectrum. As a
starting point we used the C and N abundances presented by Briley
(1997) who found a bimodal CN-distribution in 47 Tuc using a
sample of 109 RG stars. We present oxygen abundances for 103 stars
that were not affected by sky contamination. An example of
spectrum synthesis around the [O I] line is shown in Fig. \ref{fig_synth}. Altering the CN abundance by $\pm$ 0.50 dex results in a minor change on the oxygen abundance ($\mp$ 0.01 dex) as shown in Fig. \ref{fig_CNSc}a. Another source of uncertainty in the oxygen abundance is introduced by blending with Sc, therefore, the Sc abundance was adjusted for the synthesis to provide a good fit to the line profile as shown in Fig. \ref{fig_CNSc}b. 

The sodium abundance was determined using the $\lambda\lambda$  6154, 6160 \AA\ Na doublet for 138 members. Spectrum synthesis is required because
the Na $\lambda$6160 \AA\ is blended with Ca and molecular features. The line list was
the same as that used in Johnson \& Pilachowski (2012) covering
the spectral region 6150-6170 \AA. In cool, metal-poor giants the formation of Na I lines is known to exhibit departures from
local thermodynamic equilibrium (Gratton et al. 1999, Mashonkina
et al. 2000). According to Lind et al. (2011) these NLTE corrections are negligible for giants at 47 Tuc's metallicity.

The aluminum abundance was determined using the $\lambda\lambda$ 6696, 6698 \AA\ Al doublet. While the lines of this Al I doublet are not
significantly blended with other atomic lines, they are affected
by the presence of CN, particularly since 47 Tuc is relatively
metal-rich for a GC. The line list was the same as that used in Johnson \&
Pilachowski (2012). Departures from LTE in metal-poor giants for the $\lambda\lambda$ 6696, 6698 \AA\ Al doublet are less well characterized in the literature, and
aluminum abundances in Table 2 do not include NLTE corrections. 
According to Baumuller \& Gehen (1997), aluminum non-LTE
corrections are expected to be negligible for the $\lambda\lambda$ 6696, 6698 \AA\ Al doublet in metal-rich RG stars.

La and Eu were measured using spectrum synthesis to take into
account blending with nearby CN lines, hyperfine splitting, and
isotopic splitting (Eu only). The line lists used to synthesize
the 6262 \AA\ La II and 6645 \AA\ Eu II lines were the
same as that used in Johnson \& Pilachowski (2010), which include the hyperfine structure linelists available in Lawler et al. (2001a and 2001b).

\subsection {Uncertainties}

Uncertainties in the atmospheric parameters are used as independent error sources to assess the uncertainties in the abundance ratios.
The errors in our photometric temperatures introduced by the empirical fits from Alonso et al. (1999) are 25-50 K. A change in $T_{\mbox{eff}}$ of $\pm$ 50 K typically results in an abundance difference of 0.05 dex. To explore how sensitive our estimated abundances are to the adopted surface gravity, we changed log $g$ by $\pm$0.20 cgs, which is a conservative error estimation. In fact, uncertainties in the bolometric corrections and distance modulus result in an error in our photometric surface gravities of 0.10 cgs. Given the observational scatter, changes in the slope of the derived abundance vs. line strength were not statistically significant for changes in the microturbulence velocity less than 0.25 kms$^{-1}$, which we have adopted as the uncertainty in $v_t$. Sensitivities to model atmosphere parameters
are presented in Table \ref{table_sensitivities}, for a representative
star ($T_{\mbox{eff}}=4200$ K, log $g=1.3$ cgs, [Fe/H]$=$-0.75 dex, and $v_t=$1.9 kms$^{-1}$), by changing the model atmosphere parameters
($\Delta T_{\mbox{eff}}\pm50$ K, $\Delta \log g\pm0.2$ cgs, $\Delta [M/H]\pm$0.10 dex, and $\Delta v_t\pm$0.25 kms$^{-1}$)
one at a time while keeping the other unchanged, neglecting second order dependencies among the atmospheric parameters.
 
The random errors for Fe I abundances are assumed to be the standard
error of the mean, since typically 20 (Hydra sample) or 50 (FLAMES sample) Fe I lines contributed to
our measurement. For species with fewer lines, $\sigma_{\mbox{Fe I}}/\sqrt{N_{\mbox{lines}}}$ provides a realistic estimate of
observational uncertainties. For species whose abundances were estimated using spectrum synthesis, we added to the error budget the uncertainties related to smoothing, fitting, and continuum placement. The total errors presented in Table \ref{table_sensitivities} were calculated adding in quadrature the sensitivities to model atmosphere parameters and either random errors or errors introduced by the spectral synthesis technique. Overall, uncertainties in the determined abundances are smaller than 0.20 dex.

\section{Results}

\subsection {Iron and Alpha Group Elements}

\begin{figure}
\includegraphics[width=0.5\textwidth]{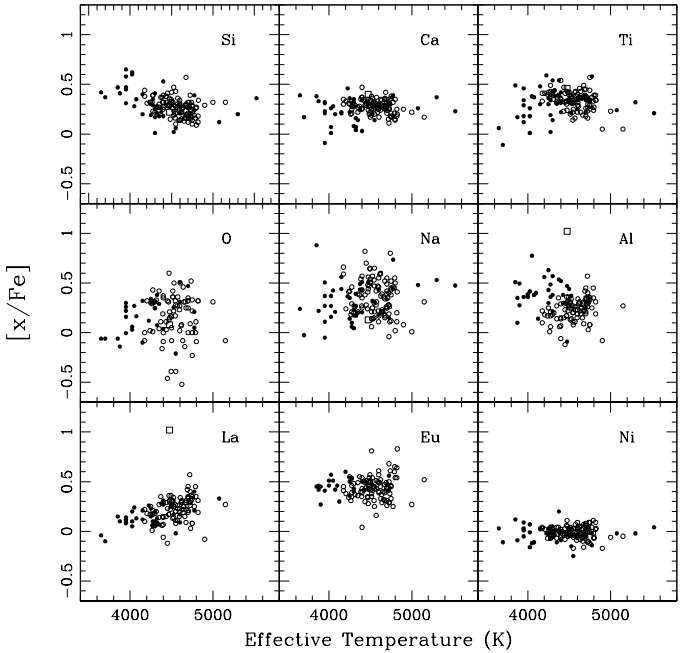} 
\caption{Derived abundance relative to Fe vs. effective temperature.  Measurements from FLAMES data are shown as open circles, and measurements from Hydra data are shown as filled circles. The open square symbol represents the La-rich star.} 
\label{fig_Abund}
\end{figure}

\begin{figure}
\includegraphics[width=0.5\textwidth]{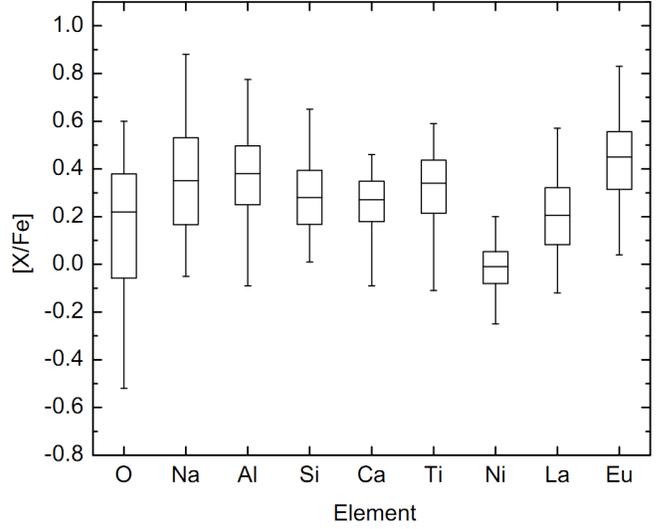} 
\caption{Element abundances and ranges measured from the FLAMES + Hydra samples.  For each element, the horizontal bar gives the median [x/Fe] ratio and the lower and upper box boundaries indicate the first and third quartiles of the data (25\% and 75\%).  The vertical lines indicate the full range of derived abundance.}
\label{fig_boxplot}
\end{figure}

In Fig. \ref{fig_boxplot}, we show a box plot with the abundance ratios for elements we have determined in our study.
Abundances ratios as a function of $T_{\mbox{eff}}$ are shown in Fig. \ref{fig_Abund}. We found [Fe/H]=$-0.79\pm0.09$ dex and
[Ni/Fe]=$-0.01 \pm 0.07$ dex, for a sample $\sim$ 164 and 140 stars respectively. Our
average [Fe/H] is consistent with other studies (Carretta et al.
2009, Koch \& McWilliam 2008, and references therein). The uncertainties due to errors in the atmospheric
parameters are comparable to the observed dispersion in both
elements, suggesting that 47 Tuc is homogeneous in both Fe and Ni.
The very small [Ni/Fe] ratio is typical of Galactic GCs and field
stars (Gratton et al., 2004).

Abundance ratios for the $\alpha$ elements are presented in Table 2 and an average [$\alpha$/Fe]=$+0.29 \pm 0.06$ dex was found. This result is in agreement within the errors with previous studies ([$\alpha$/Fe]$\sim+$0.4 dex e.g., Peterson et al. 1993, Carretta et al. 2004, Fulbright et al. 2007, Koch \& McWilliam 2008). Since the spread is
comparable to the measurement and model atmosphere uncertainties, we find that
$\alpha$-elements abundance ratios are homogeneous in 47 Tuc and
show the same trend as field metal-poor stars, i.e. they are enhanced.

\subsection {Light Elements O, Na, and Al}

\begin{figure}
\includegraphics[width=0.5\textwidth]{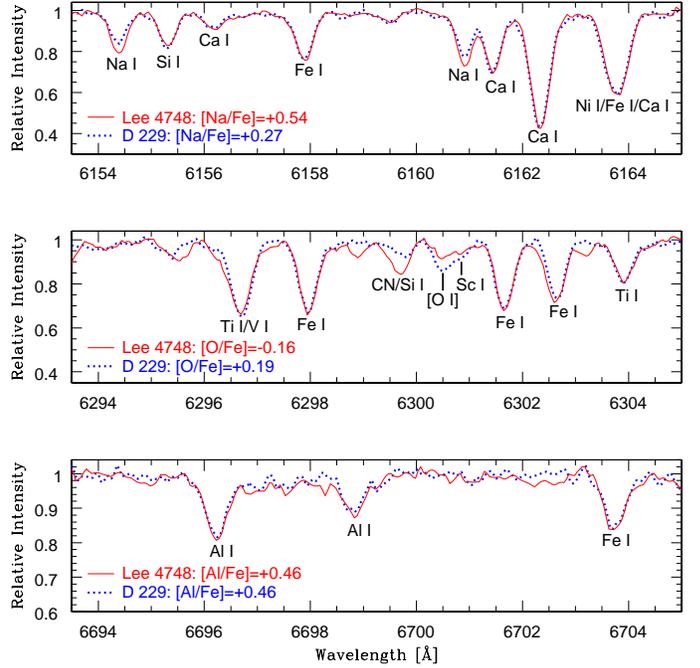} 
\caption{Stars with identical atmospheric parameters exhibiting differences in their line strengths. Interestingly, the Al content is the same for both stars, even though their O and Na abundances differ significantly.}
\label{fig_strengths}
\end{figure}

47 Tuc, like most Galactic GCs, presents variations in oxygen, sodium, and aluminum abundances. We determined O, Na, and Al abundances for a sample of 103, 138 and 129 stars, respectively and found a significant spreads in the [X/Fe] ratios for these
light elements ($\Delta (\mbox{[O/Fe]})\sim1.0$ dex, $\Delta (\mbox{[Na/Fe]})\sim0.90$ dex, and $\Delta (\mbox{[Al/Fe]})\sim0.70$ dex), which in the case of O and Al are larger than the spreads found by Carretta et al. 2009b ($\Delta (\mbox{[O/Fe]})\sim0.6$ dex and $\Delta (\mbox{[Na/Fe]})\sim0.90$ dex) and Carretta et al. 2013 ($\Delta (\mbox{[Al/Fe]})\sim0.50$ dex)). Moreover, Na and O exhibit the largest interquartile range of values of all the elements we measured in 47 Tuc. 47 Tuc exhibits the Na-O anticorrelation characteristic of many GCs (Carretta et al. 2009), shown in Fig. \ref{fig_NaO}. For this particular cluster three different populations are identified. These light element variations are not a consequence of NLTE effects, since stars with identical atmospheric parameters show different line strengths, as shown in Fig. \ref{fig_strengths}. Interestingly, the Al abundance ratios can be identical for stars that exhibit different Na abundances, suggesting that at 47 Tuc's metallicity [Al/Fe] does not strongly discriminate different populations, as shown in Figures \ref{fig_strengths} and \ref{fig_AlNa}.

On average aluminum is enhanced ([Al/Fe]=+0.38 $\pm$0.12) in our sample of 129 stars. A spectral region around the $\lambda\lambda$6696 \& 6698 \AA\ Al doublet and the spectra of two stars with almost identical stellar model atmospheres is shown in Fig. \ref{fig_strengths}. Recently, Carretta
et al. (2013) presented measurements for Al, Mg, and Si. A common sample of 20 stars indicates that our
[Al/Fe] ratios are systematically lower than their
values by 0.13 dex. The average differences in the sense our measurements minus Carretta's are $\Delta(T_{\mbox{eff}})=86$ K ($\sigma=72$ K), $\Delta(\log g)=0.11$ cgs ($\sigma=0.30$ cgs), $\Delta([Fe/H])=-0.02$ dex ($\sigma=0.10$ dex), and $\Delta(v_t)=0.26$ km s$^{-1}$ ($\sigma=0.21$ km s$^{-1}$). We found that differences in effective temperature and microturbulence velocities can be responsible for the [Al/Fe] offset in some of the stars in common.

\subsection {Neutron Capture Elements}

The star-to-star dispersion in the neutron-capture elements La and Eu is smaller than that of the lighter elements O, Na, and Al in 47 Tuc, with [La/Fe]=+0.20$\pm$0.12 dex and [Eu/Fe]=+0.44$\pm$0.11 dex, for a sample of 123 and 124 members respectively.
These abundance ratios are consistent with the range of published values in globular
clusters (e.g. Sakari et al. 2013, Worley et al. 2009, and references therein), as well as with
the abundance ratios of halo stars of similar metallicity (Gratton et al. 2004).
In the Sun, La and Eu are made primarily by the s-process and r-process, respectively,
with log A(La/Eu)=2.23 for the s-process component and log A(La/Eu)=0.09 for the
r-process component (Sneden et al. 2008). For 47 Tuc giants, we find an average value of
log A(La/Eu)=0.45, suggesting that 47 Tuc's heavy element abundances are
dominated by r-process production, with limited contribution from the s-process. Furthermore, low- and intermediate-mass AGB stars are considered to be the primary production sites for La (e.g., Busso et al. 1999; Herwig 2005; Straniero et al. 2006) and trace a longer enrichment timescale than the core-collapse SNe responsible for the enrichment of alpha-elements.  In contrast, Eu production is tied to short enrichment timescales and helps explain the variable [Eu/Fe] ratios detected in low-metallicity Galactic stars (e.g., see review by Sneden et al. 2008 and references therein).  For instance, Eu production is linked to both neutron star mergers (Freiburghaus et al. 1999; Goriely et al. 2011; Korobkin et al. 2012) and Type II SNe (Lattimer et al. 1997; Truran 1981; Mathews \& Cowan 1990; Takahashi et al. 1994; Woosley et al. 1994; Freiburghaus et al. 1999; Truran et al. 2002; Arnould et al. 2007).

The star Lee 4710 appears to be La-rich at [La/Fe]=+1.02 and [La/Eu]=+0.52. Lee 4710 may have been part of a binary mass transfer with an earlier, more massive AGB companion (Busso et al. 1999). The small
number of La-rich stars found is consistent with 47 Tuc's binary
fraction ($\leq$ 2 \%) measured by Milone et al. (2012b) using HST
data of the Globular Cluster Treasury project. Since we only have a single epoch of observation
for this star, radial velocity variations can not be estimated. Additionally, even if this La-rich star might have been in a binary system, two-body interactions with cluster stars might have made the system unbound. 

\subsection {Trends with Effective Temperature}

Si and La exhibit a trend with T$_{\mbox{eff}}$ which is shown in Fig. \ref{fig_Abund}. A linear regression of [Si I/Fe] and [La II/Fe] with T$_{\mbox{eff}}$ resulted in a correlation value of 0.44 and 0.41, respectively, which indicates a medium correlation of these abundance ratios with T$_{\mbox{eff}}$. Furthermore, the two-tailed probability (p-value) obtained to gauge the statistical significance of the linear fit is $3.00\times10^{-8}$ and $1.15\times10^{-6}$, for [Si I/Fe] and [La II/Fe] respectively. The null hypothesis that there is no linear trend of these abundance ratios with T$_{\mbox{eff}}$ is rejected since the p-values obtained are smaller than 0.05. The larger observational scatter in the [La II/Fe] ratio detected at higher temperatures is a result of the lower S/N of these fainter stars. Additionally, another consequence related to the lower S/N is that stars with intrinsically weak La lines are less likely to be included in the analysis making the slope of the [La II/Fe]-T$_{\mbox{eff}}$ relation appear steeper. Interestingly, there is a hint of a [La II/Fe] correlation with T$_{\mbox{eff}}$ for M3 and M13 in the results presented by Sneden et al (2004) in their Fig. 3. The origin of the Si-T$_{\mbox{eff}}$ trend is unknown and the high-excitation potential lines used for the analysis might not be reliable tracers of the actual Si abundance. The resolution of this long-standing problem in the abundance of Si is beyond of the scope of this project.

\section{Discussion}

\subsection{Na-O Anticorrelation}

Few chemical studies of 47 Tuc have been performed using samples of $\sim$100 or more stars. For instance, Carretta et al. (2009a) showed that the Na-O anticorrelation is a shared characteristic of all GCs, including 47 Tuc, and identified three subpopulations of primordial (P), intermediate (I), and extreme (E) stars. The spreads in O and Na are not caused by in situ nucleosynthesis, since a Na-O anti-correlation is also found in low-mass turn-off and early subgiants stars in 47 Tuc (Carretta et al. 2004), which do not have convective envelopes reaching temperatures high enough to deplete O and enhance Na. To classify 47 Tuc giants into the primordial, intermediate, and extreme populations identified by Carretta et al. (2009b), we adopted the criteria used by those authors.  The primordial population is defined as those stars with Na abundances comparable to field stars at similar metallicity; at the metallicity of 47 Tuc, this corresponds [Na/Fe]$_{\mbox{NLTE}}\leq+$0.4 dex.  Stars with [Na/Fe]$_{\mbox{NLTE}}>$0.4 dex constitute the second generation.  Additionally, Carretta et al. defined an extended second generation found in some massive GCs; this "extreme" second generation was defined as stars with  high Na and [O/Na]$_{\mbox{NLTE}}\leq$--0.90 dex.  After removing the NLTE correction used by Carretta et al.\footnote{The NLTE corrections (Gratton et al. (1999)) used by Carretta et al. are positive 
and range from +0.05 to +1.4 dex at 47 Tuc's metallicity. Therefore, 
to avoid systematic offsets introduced by using different NLTE corrections,
we have eliminated NLTE corrections from Carretta's sample in order to place their sample into our baseline.} we adopted [Na/Fe]=+0.3 dex to separate the primordial (first) and intermediate (second) generations, and [O/Na]=--0.85 dex to split the second generation into an intermediate and an extreme population. 
Our samples included several giants with lower [O/Fe] ratios than were found in Carretta et al.'s (2009b) sample.  Thus, our Na-O anti-correlation data, shown in Fig. \ref{fig_NaO}, extends in [O/Fe] down to --0.5 dex, indicating the presence of all three populations defined by Carretta et al. (2009b). A histogram of the [Na/O] abundance ratio (Fig. \ref{fig_histo}) shows two peaks with an extended tail toward high Na abundance supporting the presence of three populations. 47 Tuc is one of the few GCs in addition to Omega Cen, M13, NGC 2808 and NGC 6752, that presents more than two populations.

According to the models of Ventura and D'Antona (2008), an oxygen depletion of only $\sim0.4$ dex is expected in metal-rich AGB stars (Z=0.004) as a consequence of hot bottom burning, which makes challenging to explain the existence of third generation stars in 47 Tuc with [O/Fe]$\sim$--0.5. Interestingly, Ventura and D'Antona (2007) found that He-enhanced, metal-poor ([Fe/H]=-1.5) stars formed with [O/Fe]=--0.2 can become extreme O-poor stars ([O/Fe]$\leq$--0.4 dex) if deep mixing takes place. For instance, several authors (e.g., Kraft 1992, 1997; Pilachowski et al. 1996; Cavallo \& Nagar 2000; Sneden et al. 2004; Cohen \& Mel\'endez 2005; Johnson et al. 2005; Johnson \& Pilachowski 2012) found extreme O-poor stars in the metal-poor GC M13 and suggested that this can be explained by deep mixing. These authors found that the extreme O-depleted stars are found only above the RGB bump and their [O/Fe] ratio decreases with increasing luminosity, consistent with the deep mixing scenario. In our sample, 11 stars were found to belong to the extreme population. Even though all the stars in our sample are above the RGB bump, we do not find a strong trend of [O/Fe] versus $T_{\mbox{eff}}$. This could be a result of undersampling the cooler/more luminous stars. While extreme population, O-depleted stars are absent in the Carretta et al. (2004) sample of TO and SGB stars on 47 Tuc, nevertheless, their absence might be a consequence of a small sample statistics.  Unfortunately, simulation of deep mixing at the metallicity of 47 Tuc are unavailable to help interpret our measurements. Nonetheless, oxygen depletion by deep mixing in 47 Tuc is expected to be mild for high metallicity clusters (Charbonnel et al. 1998).

Fluorine abundances have been determined in a few GCs, such as Omega Cen, M4, NGC 6712, M22, and 47 Tuc (Cunha et al. 2003; Smith et al. 2005; Yong et al. 2008; Alves-Brito et al. 2012; Laverny \& Recio-Blanco 2013), and although F measurements are available for only small samples of stars, they exhibit star-to-star variations within the GCs. While the data available for some GCs  suggest a Na-F anticorrelation and O-F correlation, indicating that O and F are destroyed at the high temperatures needed to produce Na, de Laverny \& Recio-Blanco found a possible Na-F correlation in 47 Tuc. While observations of fluorine variations in globular clusters are intriguing, more observations are needed to quantify the variations and to clarify the role of proton-capture nucleosynthesis in globular cluster chemical evolution.

\subsection{Population Fractions and Radial Distributions}

The cumulative radial distributions of the stellar populations in our Hydra and FLAMES samples, shown in Fig. \ref{fig_cum}a, suggest that the intermediate population (green curve) is more centrally concentrated than the primordial population (blue curve).  This result is consistent with theoretical models for GC formation (D'Ercole et al. 2008) that predict that a cooling flow of low-velocity gas from AGB stars accumulates gas in the innermost regions of the cluster where a second generation of stars forms. Furthermore, VandenBerg et al. (2013) indicated that 47 Tuc is one of the massive and concentrated Galactic GCs that does not develop steady-state winds and therefore can retain the gas expelled by its stars. A Kolmogorov-Smirnov (KS) statistical test was used to assess whether the cumulative radial distributions of the different generations are drawn from the same parent distribution \footnote{We adopt a threshold of P=0.05 for the rejection of the null hypothesis that the samples are drawn from the same population.}. A P-value=0.02 allows to reject our null hypothesis. This result is in contradiction to that of Carretta et al. (2009b), who did not find a statistically significant difference in the cumulative distributions of primordial and intermediate generations.  The combination of the Hydra and FLAMES data together provides larger samples in both the outer regions of the cluster (benefiting from the wide field of view of Hydra) an the inner regions of the cluster (benefiting from the closer spatial sampling of FLAMES).  Our result is in agreement, however, with the photometric approach for tracing multiple populations (Milone et al. 2012).

After we removed the NLTE correction from the Carretta et al. Na abundances, we removed an additional systematic offset of -0.05 dex determined from a common subsample of 19 stars.  To add the Carretta et al. stars with oxygen abundances, we applied a systematic offset of +0.06 dex determined from a common subsample of 11 stars.  Finally, we combined the samples and determined population membership for all stars from the Na-O plane.  From the larger, combined sample a more significant difference between the radial distributions of the primordial and secondary populations was found, as shown in Fig. \ref{fig_cum}b, with a P-value=$2.3\times10^{-4}$.  Even though the combined sample of 191 stars with both Na and O measurements is smaller than the large sample used by Milone et al. (2012), the spectroscopic data clearly demonstrate that the second generation is more centrally concentrated than the first generation. 
The fraction of first-to-second generation stars N(I+E)/NP can be used to estimate the mass of the progenitor structure (Valcarce \& Catelan, 2011) or to constrain the long-term dynamical evolution of the cluster (Vesperini et al., 2013).  Vesperini et al. found that at any given time in the history of a cluster, the global fraction N(I+E)/NP matches that measured locally at a distance from the center of the cluster between one and two half-mass radii ($r_h$). For our combined sample, the global value for N(I+E)/NP (3.4) matches the local value at $\sim1.5r_h$, in agreement with predictions from Vesperini et al.'s N-body simulations. Additionally, the global fraction of second generation stars is expected to increase over time, since more spatially extended first generation stars are more likely to be lost from the cluster (D'Ercole et al. 2008 and Vesperini et al. 2013). 

We compare the fraction of second generation stars as a function of distance from cluster center in units of the half mass radius in Fig. \ref{fig_popratio}, which also includes measurements of the fraction of second generation stars from the Milone et al. sample.  The global value of N(I+E)/(N(I+E+P)=0.68 is equal to the local value at a radius of $\sim1.5r_h$.  Within the uncertainties of both studies, the fraction of second generation stars traces a similar profile.

Vesperini et al. (2013) also determined that multiple populations in GCs become well mixed after more than 60\% of the initial cluster mass is lost through two-body interactions (see their Fig. 4). Furthermore, Giersz \& Heggie (2011) estimated using Monte Carlo simulations that 47 Tuc has lost only $\sim45$\% of its initial mass. Thus, we would expect second generation stars still to be centrally concentrated, consistent with the observed central concentration of O-poor/Na-rich stars.

\subsection{The Aluminum Abundance in 47 Tuc}

Aluminum is the most massive light element that shows significant star-to-star variations in typical GCs and its production in significant quantities requires temperatures $\geq 70$ MK (e.g. Langer et al. 1997; Prantzos et al 2007). Since the convective envelopes of metal-rich low-mass giants do not reach these high temperatures (Lager et al. 1997, Denissenkov \& Weiss 1996, and references therein), aluminum can be used to constrain the nature of the polluters responsible for the light element abundance variations seen in GCs. 
Several studies have included analyses of the abundance of
aluminum in 47 Tuc, with most authors noting a modest enhancement of typically [Al/Fe] $\sim +0.3$ dex but little
scatter (Brown \& Wallerstein 1992, Norris \& Da Costa 1995, Carretta et al. 2004, Alves-Brito
et al. 2005, Koch \& McWilliam 2008, and Carretta et al. 2009a). Each of these studies included only a dozen or fewer stars, and a
larger sample of aluminum abundance determinations is needed to
understand the variation observed and any intrinsic spread. More recently, Carretta et al. (2013) found an average [Al/Fe] = +0.53 $\pm$
0.01 for a sample of 116 stars, and a range of $\sim$ 0.5 dex.
Moreover, O'Connell et al. (2011) noted that star-to-star variations in the
[Al/Fe] ratio within a cluster depends on metallicity.  For
clusters more metal poor than [Fe/H] $\leq$ --1.2, the spread can be as large as $\sim 1.5$ dex (see, for example, [Al/Fe] for M13 in Johnson et al.
2005, Sneden et al. 2004, and Cohen \& Mel\'endez 2005), while for more metal-rich clusters, the dispersion from
star to star in the [Al/Fe] ratio is smaller than $\sim 0.4$ dex (M107 in O'Connell et al.
2011).  

Our measured [Al/Fe] ratios are presented in a box plot in Fig. \ref{fig_boxplot}. The median abundance ratio is represented by a central horizontal line, the first and third quartiles correspond to the bottom and top box boundaries, and the full range of abundance ratios is specified by a vertical line. Comparing our results for the light elements O, Na, and Al, we first notice that the interquartile range of values is $\sim$ two times smaller for Al than for Na. This result is consistent with the simulations of Ventura \&  D'Antona (2008), hereafter VD08, which indicate that at 47 Tuc's metallicity, first generation intermediate mass ($\sim$4-8 M$_{\odot}$) AGB stars are expected to synthesize Al in smaller quantities, while producing Na in greater quantities compared to more metal-poor AGB stars. Among other physical parameters, VD08 use the mass loss treatment given by Blocker (1995), which finds greater mass loss than other published descriptions, resulting in a shorter AGB lifetime, i.e. diminishing the number of third dredge up episodes. Additionally, VD08 found that the temperature at the bottom of the convective envelope decreases with metallicity, such that only more massive AGB stars will reach the high temperatures required to convert Mg into Al. The more massive AGB stars also have shorter lifetimes limiting the Mg destruction and Al production. Thus, both the smaller number of third dredge up episodes due to shorter lifetimes and lower temperatures in the convective envelope of metal-rich AGB stars explain the lower aluminum enhancement measured in 47 Tuc and other metal-rich GCs. 

Moreover, VD08's simulations indicate that the Al spread seen in Galactic Galactic GCs can be fit by the Al-yields of AGB stars with a variety of masses. For instance, the Al spread in GCs decrease as metallicity increases, as shown in Fig. \ref{fig_boxplotAl}. This observational trend is also predicted by VD08's AGB models with masses between 3-6 $M_{\odot}$, which tend to converge towards lower Al-enhancements at higher metallicities, and differ at most by [Al/Fe]$\sim$0.4 dex, as can also be seen within the multiple populations in Omega Cen (see Fig. 22 in Johnson \& Pilachowski 2010). Thus, the metallicity of the polluters plays an important role constraining not only the degree of Al enhancement but also the spread of Al within a GC. Furthermore, the Al yields of VD08 predict a range of values of $0.3\leq$[Al/Fe]$\leq0.7$, which is consistent with the average [Al/Fe]=0.38 dex we found in our sample.

Metal-poor GCs such as M13 show an Al-Na correlation which was also detected by Carretta et al. (2013) in 47 Tuc (Fig. \ref{fig_AlNa}b). However, this correlation is not present in our sample of red giants in 47 Tuc (Fig. \ref{fig_AlNa}a). Since CN lines were not included in their line list we follow this approach and performed synthesis for a sample of 10 stars covering a Na range of 0.7 dex. We found that the Al abundances are 0.03-0.06 dex higher when the CN lines are removed from the line list. This offset is too small to account for the lack of Al-Na correlation in our data. Interestingly, M71 with a metallicity similar to 47 Tuc exhibits similar star-to-star variations in light-element abundances, including a Na-O anticorrelation, a bimodal CN distribution, and a moderate Al enhancement with a small spread $\langle$[Al/Fe]=$+0.24\pm0.10\rangle$. The Al-Na correlation in M71, however, relies on a single star (Ramirez \& Cohen 2004). Furthermore, RGB stars in Omega Cen with similar metallicity to 47 Tuc giants do not exhibit an Al-Na correlation (Johnson \& Pilachowski 2010). To assess whether the lack of an Al-Na anticorrelation is a common characteristic in metal-rich GCs, large samples of stars with measured Al and Na abundances are needed in other more metal-rich GCs.

\section{Summary}
The motivation for pursuing this project comes from the discovery that most GCs host multiple populations (Carretta et al. 2009a) and the properties of their generations can be used to test models of GC formation or models that describe the nature of the polluters. A direct method to identify the different generations in a GC relies on the chemical content of light-elements. Therefore, we have used abundance ratios of O, Na, and Al to characterize 47 Tuc's multiple populations. Additionally, we have measure abundance ratios for alpha, Fe peak, and neutron capture group elements, which can be used to compare 47 Tuc with other Galactic GCs and halo stars. 

We summarize our results as the following.

1. Chemical abundances for O, Na, Al, Si, Ca, Ti, Fe, Ni, La, and Eu
were determined for 164 stars in 47 Tuc using either spectral
synthesis or EW analysis. Abundance ratios for the $\alpha$ elements, Si, Ca, and Ti, and the n-capture process
elements, Eu and La, indicate that the composition of the first generation stars is dominated by Type II SNe ([$\alpha$/Fe]$\approx$0.3 and
[Eu/La]=+0.24). 

2. Star-to-star variations were detected only for O, Na, and Al. We found that the Na-O anticorrelation is more extended that previously reported reaching [O/Fe] ratios as low as $\sim$--0.5 dex, as found typically in the more massive Galactic GCs. Three groups can be identified in the Na-O plane indicating that 47 Tuc hosts multiple populations, with younger stars being enhanced in Na and depleted in O.

3. For the first time Na and O measurements indicate that the intermediate generation is more
centrally concentrated than the primordial generation, which is
consistent with current hydrodynamical N-body simulations of GC
formation. A KS test indicates that the detected difference in radial distributions is real and not a
result of random fluctuations. Furthermore, the radial distributions of first and second generation stars found using Na and O are consistent with the results obtained from photometric data. Additionally, our spectroscopic measurements show that the population ratio varies with distance as predicted by N-body simulations (Vesperini et al. (2013).  

4. No Na-Al correlation is found in 47 Tuc. To assess whether this is characteristic of more metal-rich GCs, large samples in several clusters are needed. Moreover, we found that the different populations can not be identified using their [Al/Fe] ratio, since stars with identical atmospheric parameters and different O and Na abundances can display the same Al content. 

5. A single La-rich star (Lee 4710) was found in our sample whose anomalous La abundance could be explained by mass transfer in a close binary system. Comparing radial velocities at different epochs can reveal whether the star is currently in a binary system. A single binary system in our sample is consistent with the small binary fraction measured in this cluster. However, even if this star was originally part of a binary system, two-body interactions with other cluster stars might have made the system unbound.

\acknowledgments
MJC wish to thank Enrico Vesperini for his helpful comments and discussions. We also thank an anonymous referee for his/her thoughtful comments, which have clarified and improved this work. We are grateful to the Cerro Tololo InterAmerican Observatory for assistance in obtaining the observations. This research has made use of the NASA Astrophysics Data System Bibliographic Services. This publication makes use of data products from the Two Micron All Sky Survey, which is a joint project of the University of Massachusetts and the Infrared Processing and Analysis Center/California Institute of Technology, funded by the National Aeronautics and Space Administration and the National Science Foundation.  CIJ gratefully acknowledges the support of the National Science Foundation under award AST-1003201. CAP acknowledges the generosity of the Kirkwood Research Fund at Indiana University.

\begin{figure}
\includegraphics[width=0.5\textwidth]{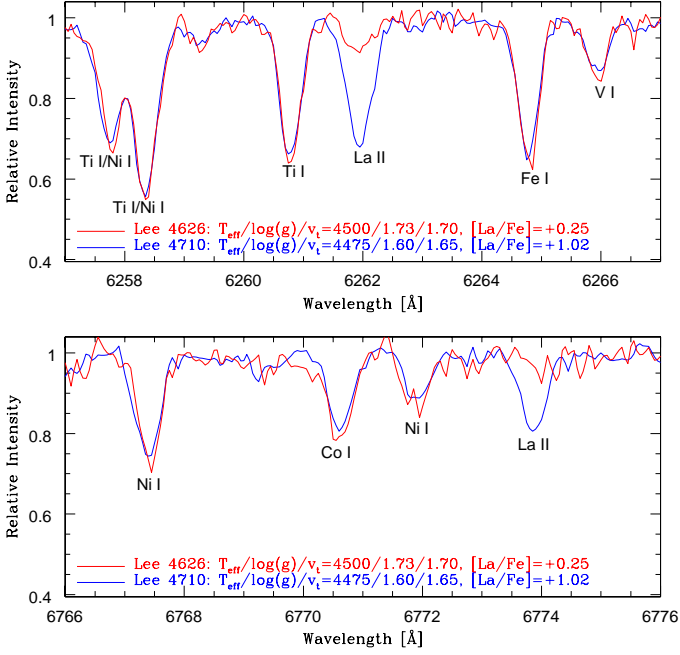} 
\caption{A spectrum of the La-rich giant Lee 4710 vs. Lee 4626, a giant of similar atmospheric parameters with normal La.   Identified features of Ti, V, Fe, and Ni show similar strength in the two stars, while the features of La II at 6262 and 6774 \AA\  are stronger in Lee 4710 than in Lee 4626. } 
\label{fig_binary}
\end{figure}

\begin{figure}
\includegraphics[width=0.5\textwidth]{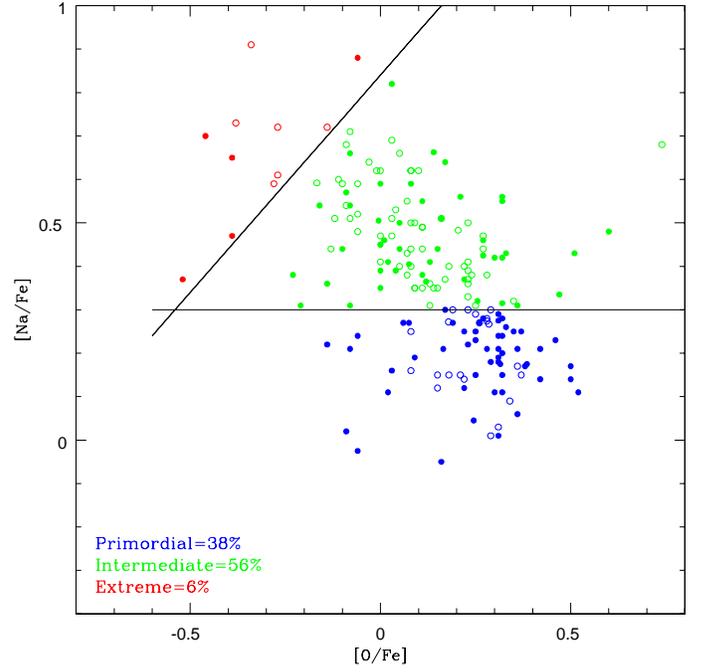} 
\caption{The Na-O anticorrelation.  Stars assigned to the primordial population are shown in blue, stars assigned to the intermediate population are shown in green, and stars assigned to the extreme population are shown in red.  Our FLAMES and Hydra samples are represented by filled circles, while from Carretta et al. (2009b) are shown with open symbols.  Stars are assigned to the primordial, intermediate, and extreme populations as defined in the text and indicated by solid lines in the figure.}
\label{fig_NaO}
\end{figure}

\begin{figure}
\includegraphics[width=0.5\textwidth]{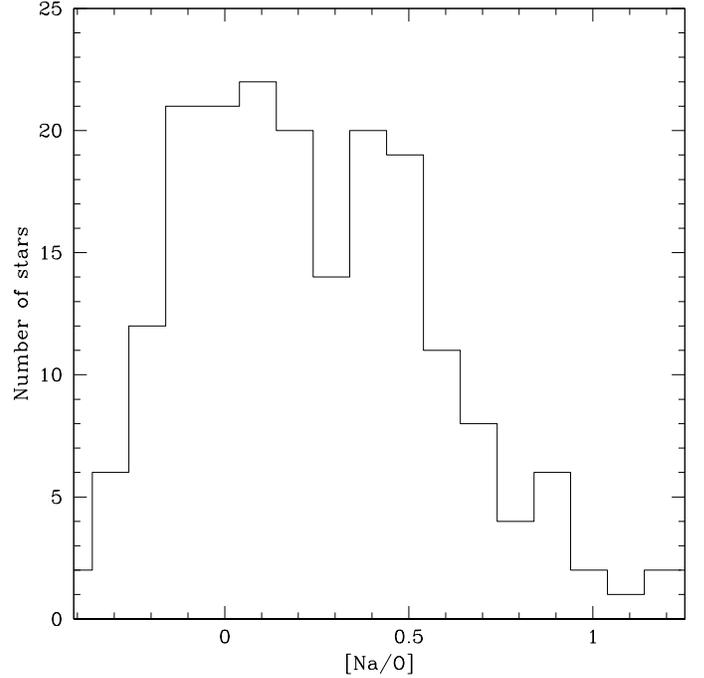} 
\caption{Histogram of the [Na/O] abundance ratio distribution showing two peaks with an extended tail.}
\label{fig_histo}
\end{figure}

\begin{figure}
\includegraphics[width=0.5\textwidth]{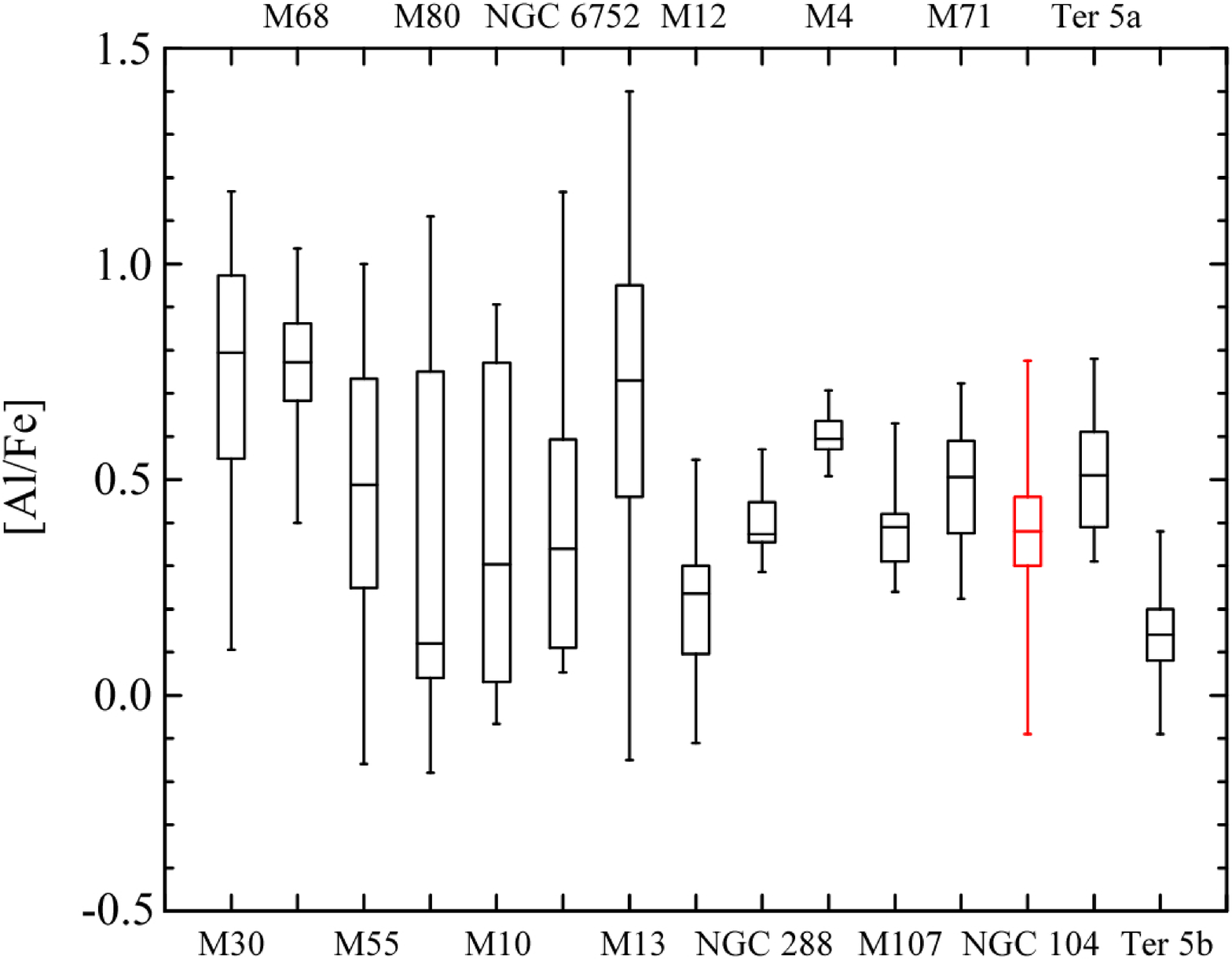} 
\caption{The [Al/Fe] abundance in 47 Tuc from the FLAMES + Hydra samples compared to other globular clusters.  For each cluster, the horizontal bar gives the median [Al/Fe] ratio and the lower and upper box boundaries indicate the first and third quartiles of the data (25\% and 75\%).  The vertical lines indicate the full range of the derived aluminum abundances in each cluster. Data sources are as follows:  M80 from Cavallo et al. (2004); M13 from Johnson et al. (2005); M30, M68, M55, M10, NGC 6752, M12, NGC 288, and M4, and M71, from Carretta et al. (2009b); M107 from O'Connell et al. (2011); Terzan 5 from Origlia et al. (2011), where a and b represents the populations with [Fe/H]=--0.25 and [Fe/H]=+0.27, respectively.}
\label{fig_boxplotAl}
\end{figure}

\begin{figure}
\includegraphics[width=0.5\textwidth]{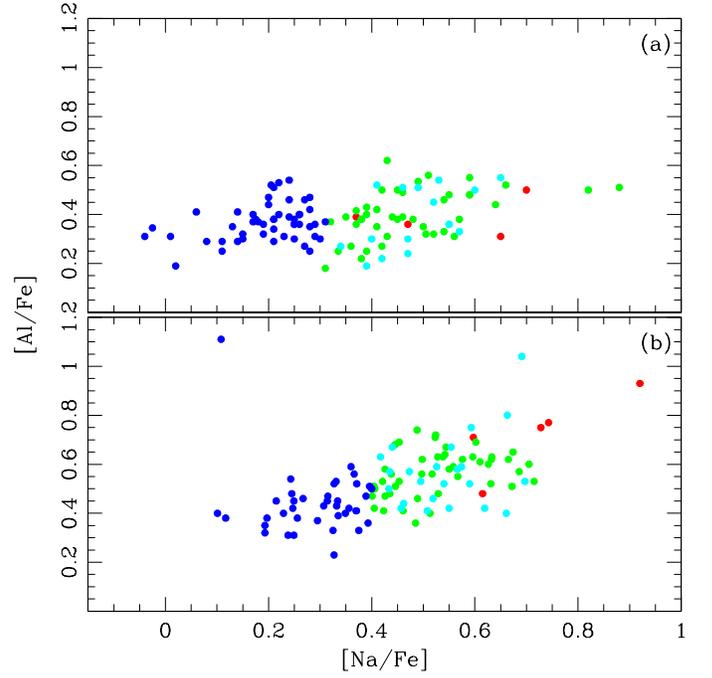} 
\caption{[Al/Fe] vs. [Na/Fe]. Stars assigned to the primordial population are shown in blue, stars assigned to the intermediate population are shown in green, and stars belonging to the extreme population are shown in red.  Stars not classified according to population are shown in cyan. The upper panel (a) shows that in our sample both the primordial and intermediate populations exhibit a similar mean aluminum abundance and dispersion. The lower panel (b) shows results from Carretta et al. 2013.} 
\label{fig_AlNa}
\end{figure}

\begin{figure}
\includegraphics[width=0.5\textwidth]{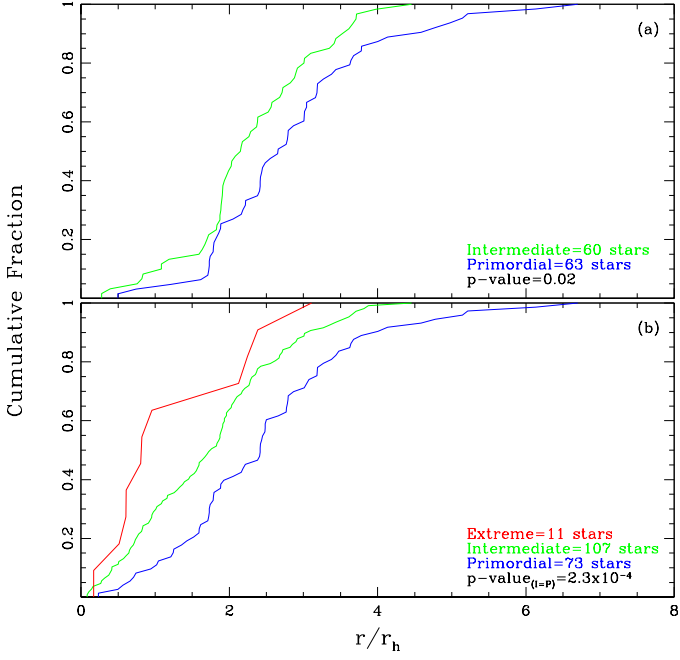}
\caption{(upper panel) Cumulative fraction of stars as a function of half-mass radius for the FLAMES and Hydra samples.  The curve for intermediate population stars is shown in green and the curve for primordial population stars is shown in blue.  The intermediate population is more centrally concentrated.  A p-value of 0.02 from the KS test allows us to reject the null hypothesis that the two samples are drawn from the same population. (lower panel) Cumulative fraction of stars as a function of half-mass radius for our data plus the Carretta et al. (2009b) sample.  The cumulative fraction for the combined sample is shown with a black curve, and the extreme population is shown as a red curve.  As in Figure 9, the curve for intermediate population stars is shown in green and the curve for primordial population stars is shown in blue.  The intermediate population is more centrally concentrated.  A p-value of $2.3\times10^{-4}$  from the KS test allows us to reject more strongly the null hypothesis that the samples are drawn from the same population.}
\label{fig_cum}
\end{figure}

\begin{figure}
\includegraphics[width=0.5\textwidth]{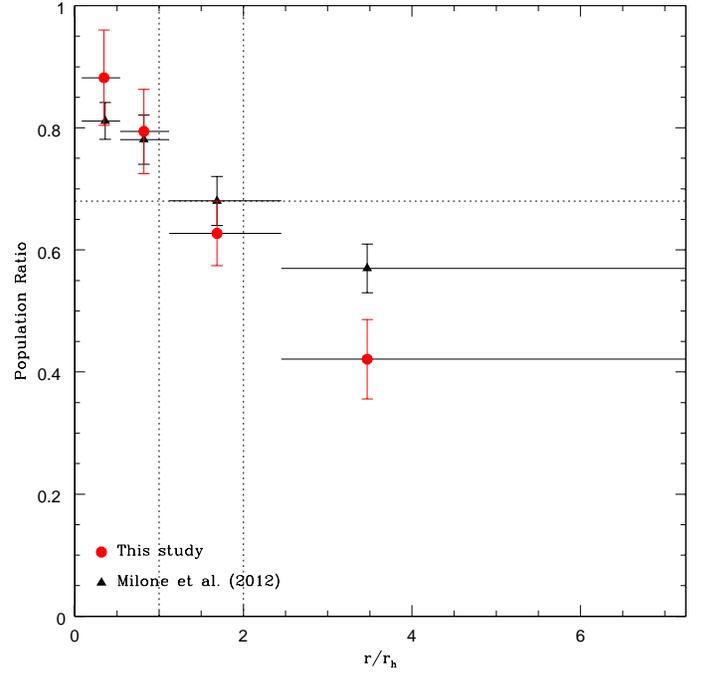} 
\caption{The fraction of subsequent (intermediate plus extreme) generation stars in 47 Tucanae vs. cluster half mass radius.  Data from the combined FLAMES, Hydra, and Carretta et al. (2009b) are shown in red, while population fractions from Milone et al. (2012) are shown in black.  The intermediate plus extreme populations dominate in the central regions, while the primordial population dominates beyond about 1.5 half-mass radii from the cluster center.}
\label{fig_popratio}
\end{figure}

\clearpage

\clearpage

\begin{table}
\begin{center}
\begin{tabular}{ccc}
\hline\hline
UT Date & Exp. Time & Wavelength  \\
\& Time & (s) & ($\AA$) \\
\hline
18 July 2003 06:53& 3600 & 6490-6800 \\
18 July 2003 07:54& 3600 & 6490-6800 \\
18 July 2003 09:00& 3600 & 6490-6800 \\
19 July 2003 06:41& 3600 & 6490-6800 \\
19 July 2003 07:50& 3600 & 6490-6800 \\
19 July 2003 08:52& 3600 & 6490-6800 \\
19 July 2003 10:00& 1800 & 6490-6800 \\
23 Aug 2010 06:32 & 2700 & 6100-6400 \\
23 Aug 2010 07:21 & 2700 & 6100-6400 \\
23 Aug 2010 08:09 & 2700 & 6100-6400 \\
23 Aug 2010 08:58 & 2700 & 6100-6400 \\
26 Nov 2011 00:26 & 2100 & 6110-6396 \\
26 Nov 2011 01:05 & 1020 & 6300-6390 \\
26 Nov 2011 01:25 & 960  & 6600-6989 \\
\hline\hline
\end{tabular} 
\caption{Log of 47 Tucanae Observations}
\label{table_log}
\end{center} 
\end{table}

\begin{sidewaystable}
\begin{tabular}{lrrrcrrrrrrrrrcr}
\hline\hline
&T$_{\text{eff}}$ & & & $v_t$ & & & & & & & & & & & r \\
Star & (K) & $\log g$ & [Fe/H] & (km s\textsuperscript{-1}) & [O/Fe] & [Na/Fe] & [Al/Fe] & [Si/Fe] & [Ca/Fe]
& [Ti/Fe] & [Ni/Fe] & [La/Fe] & [Eu/Fe] & Pop & (arcmin) \\
\hline
Hydra\\
1628 & 4350 & 1.5 & -0.73 & 1.6 & 0.29 & 0.2 &    & 0.18 & 0.14 & 0.31 & -0.11 & 0.07 &    & P & 9.53\\
1646 & 4150 & 1.2 & -0.72 & 1.75 & -0.1 & 0.44 &    & 0.2 & 0.21 & 0.3 & 0.02 & 0.12 &    & I & 5.46\\
1669 & 4775 & 1.95 & -0.79 & 2.35 & 0.21 & 0.735 &    & 0.39 & 0.39 & 0.58 & -0.02 & 0.4 &    & I & 5.99\\
1735 & 4525 & 1.7 & -0.73 & 1.75 & 0.255 & 0.32 & 0.37 & 0.02 & 0.2 & 0.21 & -0.15 & 0.17 & 0.47 & I & 7.55\\
1747 & 4025 & 1.05 & -0.81 & 2.0 &  0.03  & 0.16 &    & 0.60 & 0.07 & 0.01 & -0.01 &  0.10  &    & P & 6.85\\
\\
FLAMES\\
1628 &   4400    &   1.23    &   -1.00   &   1.80    &   0.35    &   0.43    &   0.62    &   0.27    &   0.47    &   0.54    &   -0.01   &   0.14    &   0.39    &    I  &   9.55    \\
1669   &   4615    &   2.10    &   -0.72   &   1.60    &   0.07    &   0.59    &   0.55    &   0.22    &   0.29    &   0.28    &   -0.03   &   0.29    &   0.49    &    I  &   5.98    \\
1703    &   4575    &   2.00    &   -0.79   &   1.55    &   0.36    &   0.31    &   0.37    &   0.36    &   0.37    &   0.43    &   0.07    &   0.23    &   0.43    &    P  &   8.09    \\
1714    &   4450    &   1.75    &   -0.82   &   1.65    &   0.32    &   0.28    &   0.47    &   0.29    &   0.29    &   0.41    &   0.03    &   0.23    &   0.51    &    P  &   7.77    \\
1716    &   4700    &   2.47    &   -0.66   &   1.35    &   0.25    &   0.25    &   0.36    &   0.21    &   0.21    &   0.18    &   -0.05   &   0.35    &    &    P  &   7.72    \\
\hline\hline
\end{tabular} 
\caption{Atmospheric Parameters and Abundances.}
\label{tabshort}
\end{sidewaystable}

\begin{table}
\begin{center}
\begin{tabular}{lllllllll}
\hline\hline
Ion & $T_{\mbox{eff}}\pm50$ & $\log g\pm0.20$ & [M/H]$\pm$0.10 & $v_t\pm$0.25 & No. & $\sigma_{\mbox{obs.}}$ & $\sigma_{\mbox{total}}$ & offset \\
    & (K) & (cgs) & (dex) & (km s$^{-1}$) & lines & (dex) & (dex) & (dex) \\
\hline
Fe I        &   $\mp    0.01    $   &   $\mp    0.03    $   &   $\pm    0.02    $   &   $\pm    0.14    $   &   20,50   &   0.03,0.02   &   0.15  & --0.08      \\
Fe II       &   $\mp    0.07    $   &   $\pm    0.08    $   &   $\pm    0.05    $   &   $\mp    0.03    $   &   0,6 &  ,0.08   &  ,0.15  &   \\
O I     &   $\pm    0.03    $   &   $\pm    0.09    $   &   $\pm    0.05    $   &   $\pm    0.05    $   &   1,1 &   0.10        &   0.15    & --0.07    \\
Na I        &   $\pm    0.03    $   &   $\mp    0.05    $   &   $\pm    0.05    $   &   $\mp    0.05    $   &   2,2 &   0.10        &   0.14  & +0.07   \\
Al I        &   $\pm    0.03    $   &   $\mp    0.02    $   &   $\mp    0.02    $   &   $\mp    0.05    $   &   2,2 &   0.10        &   0.12 & +0.07       \\
Si I        &   $\mp    0.04    $   &   $\pm    0.03    $   &   $\pm    0.03    $   &   $\mp    0.03    $   &   1,2 &   0.15,0.11   &   0.16,0.13 & 0.00  \\
Ca I        &   $\pm    0.07    $   &   $\mp    0.02    $   &   $\pm    0.00    $   &   $\mp    0.15    $   &   2,2 &   0.11        &   0.20 & --0.09    \\
Ti I        &   $\pm    0.10    $   &   $\pm    0.01    $   &   $\pm    0.00    $   &   $\mp    0.09    $   &   2,2 &   0.11        &   0.17  & +0.02 \\
Ni I        &   $\pm    0.00    $   &   $\pm    0.04    $   &   $\pm    0.02    $   &   $\mp    0.09    $   &   2,2 &   0.11        &   0.15 & --0.05       \\
La II       &   $\pm    0.01    $   &   $\pm    0.09    $   &   $\pm    0.06    $   &   $\pm    0.01    $   &   1,2 &   0.10,0.08   &   0.15,0.14   & +0.04     \\
Eu II       &   $\mp    0.03    $   &   $\pm    0.05    $   &   $\pm    0.05    $   &   $\mp    0.03    $   &   1,1 &   0.10        &   0.13 & +0.04     \\
\hline\hline
\end{tabular} 
\caption{Abundance Sensitivity to Model Atmosphere Parameters.}
\label{table_sensitivities}
\end{center} 
\end{table}

\end{document}